\pdfoutput=1
\documentclass[sigconf]{acmart}

\usepackage{svg}
\usepackage{colortbl}

\usepackage{pifont}
\usepackage[utf8]{inputenc} % allow utf-8 input
\usepackage{url}            % simple URL typesetting
\usepackage{amsfonts}       % blackboard math symbols
\usepackage{nicefrac}       % compact symbols for 1/2, etc.
\usepackage{enumitem}
\usepackage{rotating}
\usepackage{amsthm}

\usepackage{multirow}
\usepackage{bm}
\usepackage{algorithm}
\usepackage{algorithmic}
\colorlet{backred}{ACMRed!35}
\colorlet{backblue}{ACMLightBlue!65}
\colorlet{lightyellow}{ACMYellow!45}
\colorlet{DarkGreen}{green!55!black}

\hypersetup{
    colorlinks=true,
    linkcolor=red,
    citecolor=cyan,
    filecolor=magenta,      
    urlcolor=magenta,
    }

\usepackage{listings}

\usepackage{makecell}
\usepackage{subcaption}
\newcommand{\method}{MIME}

\AtBeginDocument{%
  }

\copyrightyear{2026}
\acmYear{2026}
\setcopyright{cc}
\setcctype{by}
\acmConference[MM '26]{Proceedings of the 34th ACM International Conference on Multimedia}{November 10--14, 2026}{Rio de Janeiro, Brazil}
\acmBooktitle{Proceedings of the 34th ACM International Conference on Multimedia (MM '26), November 10--14, 2026, Rio de Janeiro, Brazil}
\acmDOI{10.1145/3767308.3838643}
\acmISBN{979-8-4007-2213-4/2026/11}

\begin{document}

\title{Face Is Not All You Need: MIME Benchmark for Incomplete Multimodal Emotion Recognition}

\author{Yuxin Jia}
\email{jiayuxin@mails.neu.edu.cn}
\affiliation{%
  \institution{School of Computer Science and Engineering, Northeastern University}
  \city{Shenyang}
  \country{China}
}
\affiliation{
  \institution{Hubei Key Laboratory of Multimedia and Network Communication Engineering, Wuhan University}
  \city{Wuhan}
  \country{China}
}

\author{Xing Lan}
\email{lanx@cse.neu.edu.cn}
\correspondingauthor
\authornote{Corresponding author: lanx@cse.neu.edu.cn}
\affiliation{%
  \institution{School of Computer Science and Engineering, Northeastern University}
  \city{Shenyang}
  \country{China}
}
\affiliation{%
  \institution{State Key Lab. of Novel Software Technology, Nanjing 
University}
  \city{Nanjing}
  \country{China}
}

\author{Wensong Wang}
\email{20235870@stu.neu.edu.cn}
\affiliation{%
  \institution{School of Computer Science and Engineering, Northeastern University}
  \city{Shenyang}
  \country{China}
}
\affiliation{
  \institution{National Engineering Research Centre for Multimedia Software (NERCMS), Wuhan University}
  \city{Wuhan}
  \country{China}
}

% \author{Huiming Zhu}
% \affiliation{%
%   \institution{School of Computing, National University of Singapore}
%   % \city{Singapore}
%   \country{Singapore}
% }

\author{Jian Xue}
\email{xuejian@ucas.ac.cn}
\affiliation{%
  \institution{School of Engineering Science, University of Chinese Academy of Sciences}
  \city{Beijing}
  \country{China}
}

\author{Feiliang Ren}
\email{renfeiliang@cse.neu.edu.cn}
\affiliation{%
  \institution{School of Computer Science and Engineering, Northeastern University}
  \city{Shenyang}
  \country{China}
}

\author{Ke Lu}
\email{luk@ucas.ac.cn}
\affiliation{%
  \institution{School of Engineering Science, University of Chinese Academy of Sciences}
  \city{Beijing}
  \country{China}
}

\renewcommand{\shortauthors}{Yuxin Jia et al.}
\newcommand\datasetabbr{GroundLie360}
\newcommand\faketypeone{factual distortion}
\newcommand\Faketypeone{Factual distortion}
\newcommand\faketypetwo{spatial manipulation}
\newcommand\Faketypetwo{Spatial manipulation}
\newcommand\faketypethree{temporal manipulation}
\newcommand\Faketypethree{Temporal manipulation}
\newcommand\faketypefour{false speech}
\newcommand\Faketypefour{False speech}
\newcommand\faketypefive{contradictory content}
\newcommand\Faketypefive{Contradictory content}
\newcommand\faketypesix{unsupported content}
\newcommand\Faketypesix{Unsupported content}
\begin{abstract}

Multimodal Emotion Recognition (MER) is a fundamental task in multimedia understanding, where state-of-the-art methods have achieved prominent success relying on the assumption of complete, well-aligned multimodal inputs. However, real-world unconstrained scenarios often suffer from unpredictable modality degradation and information loss, leading to catastrophic performance collapse of existing MER models that heavily depend on facial cues as the core inference anchor. Worse still, there remains a critical research gap: few systematic works exist to evaluate model robustness under diverse, fine-grained incomplete information settings, as prior works are limited to coarse-grained full-modality removal and neglect interpretable analysis of emotional reasoning processes. To bridge this gap, we formally define the Incomplete Multimodal Emotion Recognition (IMER) task, which benchmarks model generalization and robustness under multi-grained modality and emotional information loss. We further construct MIME, a dedicated IMER benchmark with 2,000 video segments covering realistic fine-grained facial-detail degradation, severe facial degradation, and extreme full-modality loss, surpassing the limited coarse-grained settings of prior works. We also propose the Chain of Emotion (CoE) analysis paradigm with tailored evaluation metrics to dissect the fine-grained emotional cue perception and reasoning process of Multimodal Large Language Models (MLLMs). Extensive experiments on seven mainstream MLLMs reveal that body language and scene context act as critical compensatory mechanisms under facial degradation, while a pronounced gap exists between models' basic visual perception ability and high-level cognitive reasoning. Our work provides a focused evaluation platform and actionable insights for advancing robust multimodal emotion recognition. Our dataset and code are public at https://yuxinokk.github.io/MIME.

\end{abstract}

% \vspace{-8mm}


\maketitle
\section{Introduction}
Multimodal Emotion Recognition (MER) is an important task in multimedia, aiming to infer human emotional states from multimodal information \cite{kalateh2024systematic, wu-etal-2025-multimodal, hu2024recent, LI2026114127}. 
Recent advancements in MER ~\cite{NAACL2025,llava,mer2025} have demonstrated remarkable success by effectively fusing diverse multimodal information, such as facial dynamics, vocal nuances, body language, and contextual visual cues, etc, which usually rely on the assumption of completed and well-aligned multimodal inputs.

However, in unconstrained real-world deployment scenarios, the acquired multimodal data often suffers from unpredictable degradation and modality missing due to various practical factors~\cite{guo2024multimodal, liu2024contrastive, sun2025enhancing, zhong2025towards}, such as severe facial occlusion, lighting fluctuations, sensor malfunctions, network packet loss, and privacy constraints. Most existing MER methods~\cite{waligora2024joint, lian2026merbench, lian2025mer, wu2025multimodal} heavily rely on facial expression features as the core and dominant cue for emotion inference, which leads to catastrophic performance degradation or even complete functional failure when the critical facial modality is corrupted, invalidated, or entirely unavailable. 

Despite the urgent demand for robust MER systems~\cite{shou2025multimodal, deng2025multiplex, jin2025rohydr} in practical scenarios, there remains a notable and unexplored research gap: few works have systematically investigated and benchmarked how current MER models perform under diverse, fine-grained, incomplete information settings. 
To bridge this gap, we formally define a novel task: \textbf{I}ncomplete \textbf{M}ultimodal \textbf{E}motion \textbf{R}ecognition (\textbf{IMER}), which focuses on evaluating the generalization ability and robustness of emotion recognition models under varying levels of modality and emotional information missing. 
Compared with the standard MER task built on ideal complete-modal assumptions, IMER can more accurately reflect the real-world performance and practical reliability of emotion recognition systems in unconstrained deployment environments.
% By systematically exploring MLLMs performance under various scenarios of modality loss, the IMER task serves as a crucial proxy for evaluating the true robustness, reliability, and fault tolerance of emotion recognition systems in practical applications.

To facilitate the proposed task, we construct \textbf{M}issing \textbf{I}nformation in \textbf{M}ultimodal \textbf{E}motion (\textbf{MIME}), a dedicated benchmark tailored for assessing model performance under diverse information-missing scenarios, which contains 2,000 video segments. 
Unlike existing IMER benchmarks that only simulate extreme, coarse-grained full-modality missing (e.g., complete absence of audio or visual streams), MIME emphasizes that real-world incompleteness is inherently multi-grained: in many practical cases, the entire visual modality is not fully lost, but the critical facial emotional evidence is only partially degraded. We therefore distinguish fine-grained facial-detail degradation from more severe facial degradation, rather than assuming that facial evidence is either fully available or entirely absent.

In addition, we design a novel analysis paradigm termed Chain of Emotion (\textbf{CoE}), to dissect the fine-grained emotional cue perception and reasoning process of Multimodal Large Language Models (MLLMs) during emotion recognition. Specifically, CoE explicitly guides models to decouple and articulate their observations across individual modalities—namely face, body, scene, and audio—prior to predicting the final emotion, moving beyond the single final accuracy metric to provide interpretable insights into model behaviors. 
Furthermore, we develop task-specific evaluation metrics for CoE, which are designed for scene understanding, emotional analysis, classification Accuracy and overall CoE quality.

Our extensive evaluation results reveal that while the face is the primary emotional indicator, body language and scene context act as essential compensatory mechanisms under facial degradation (effectively mitigating catastrophic performance collapse). 
Even the most powerful MLLMs suffer from significant performance drops when facing fine-grained facial information corruption and cross-modality joint missing scenarios, exposing severe robustness limitations of existing models in real-world settings. 
CoE evaluations expose a pronounced gap between the models' fundamental visual perception capabilities and their high-level cognitive reasoning skills. 

The main contributions of our work are summarized as follows:
\begin{itemize}
\item We formally define the IMER task, the first systematic task formulation focusing on multi-grained information missing in MER, which bridges the long-standing gap between standard ideal settings and real-world deployment demands.

\item We construct MIME, the first dedicated benchmark for IMER, equipped with the CoE analysis paradigm and task-specific evaluation metrics, providing a standardized and extensible evaluation platform for the research community.

\item We conduct comprehensive benchmark experiments on 7 mainstream open-source and closed-source MLLMs under various missing-modality conditions.
We reveal their strengths and limitations, paving the way for future research efforts to advance emotional intelligence in MLLMs

% By delivering in-depth analyses on modality reliance, compensatory mechanisms, and the gap between perception and reasoning, we provide actionable insights to guide future research on robust multimodal emotion recognition.

\end{itemize}

\section{Related Work}

\subsection{Multimodal Emotion Recognition Datasets}

The evolution of Multimodal Emotion Recognition (MER) has been fundamentally driven by large-scale datasets. Early affective computing research relied on laboratory-collected datasets like IEMOCAP \cite{IEMOCAP} and MELD \cite{MELD}. To capture more naturalistic human emotions, recent benchmarks have shifted toward \textit{in-the-wild} scenarios. For instance, CAER \cite{CAER} pioneered context-aware emotion recognition by emphasizing the joint contribution of facial expressions and background scenes. DFEW \cite{DFEW} and MAFW \cite{MAFW} provided large-scale dynamic video clips extracted from real-world media, encompassing rich audio-visual multimodal cues. Furthermore, FERV39k \cite{FERV39k} contributed a massive collection of unconstrained, silent video clips, driving the research of visual-only emotion analysis. However, a fundamental limitation of these benchmarks is their reliance on the assumption of modality completeness. Furthermore, their annotations are strictly limited to discrete categorical labels (e.g., \textit{Happy}, \textit{Sad}). They lack the fine-grained reasoning rationales required to interpret complex emotional states\cite{sun2025enhancing, zhong2025towards, YAZICI2026200642, wu2024deep}, significantly limiting their utility in training interpretable affective models.

\subsection{Emotion Recognition under Missing Modalities}
Recent works~\cite{cvpr2024CorrKD,ACL2024,NeurIPS2025FedDISC} explore MER under missing modalities, but they are limited to a coarse-grained setting that simply removes fixed modalities from standard datasets and retains only 1-2 modalities. This paradigm cannot support evaluation under semantically specific information loss, nor can it model emotional reasoning processes or analyze the synergistic effect of effective modalities on emotion recognition~\cite{zhou2024review}.

To address this gap, we build MIME benchmark with multiple granularities for missing modalities.
Unlike previous datasets, our benchmark uniquely decouples visual elements into distinct fine-grained modalities and supports comprehensive missing-modality evaluation for robustness.

\begin{table}[t]
\centering
\caption{\textbf{Comparison of MIME with other Benchmarks related to IMER.} }
\label{tab:bench_cmp}
\resizebox{\columnwidth}{!}{%
\begin{tabular}{l c c c c}
\toprule
\textbf{Benchmark} & \textbf{Format} & \textbf{\makecell{Supp. \\ Incomplete}} & \textbf{\makecell{Incom. \\ grained}} & \textbf{\makecell{Chain of \\ Emotion}} \\
\midrule
EmotionBench \cite{huang2023emotionally} & Text & \textcolor{ACMRed}{\ding{55}} & \textcolor{ACMRed}{\ding{55}} & \textcolor{ACMRed}{\ding{55}} \\
EmoBench \cite{sabour2024emobench}       & Text & \textcolor{ACMRed}{\ding{55}} & \textcolor{ACMRed}{\ding{55}} & \textcolor{ACMRed}{\ding{55}} \\
MOSABench \cite{song2024mosabench}       & Image& \textcolor{ACMRed}{\ding{55}} & \textcolor{ACMRed}{\ding{55}} & \textcolor{ACMRed}{\ding{55}} \\
MM-InstructEval \cite{yang2025mm}        & Image& \textcolor{ACMRed}{\ding{55}} & \textcolor{ACMRed}{\ding{55}} & \textcolor{ACMRed}{\ding{55}} \\
IEMOCAP \cite{busso2008iemocap}          & Video& \textcolor{DarkGreen}{\ding{51}} & Coarse& \textcolor{ACMRed}{\ding{55}} \\
MC-EIU \cite{liu2024emotion}             & Video& \textcolor{ACMRed}{\ding{55}} & \textcolor{ACMRed}{\ding{55}} & \textcolor{ACMRed}{\ding{55}} \\
OV-MER \cite{lian2024open}               & Video& \textcolor{ACMRed}{\ding{55}} & \textcolor{ACMRed}{\ding{55}} & \textcolor{ACMRed}{\ding{55}} \\
EmoBench-M \cite{hu2025emobench}         & Video& \textcolor{ACMRed}{\ding{55}} & \textcolor{ACMRed}{\ding{55}} & \textcolor{ACMRed}{\ding{55}} \\
MER-UniBench \cite{lian2024affectgpt}    & Video& \textcolor{ACMRed}{\ding{55}} & \textcolor{ACMRed}{\ding{55}} & \textcolor{ACMRed}{\ding{55}} \\
MOSI \cite{mosi}                         & Video& \textcolor{DarkGreen}{\ding{51}} & Coarse& \textcolor{ACMRed}{\ding{55}} \\
MME-Emotion \cite{zhang2025mme}          & Video& \textcolor{ACMRed}{\ding{55}} & \textcolor{ACMRed}{\ding{55}} & \textcolor{DarkGreen}{\ding{51}} \\
\midrule
\rowcolor{black!10} \textbf{MIME (Ours)} & Video& \textcolor{DarkGreen}{\ding{51}} & Fine \& Coarse& \textcolor{DarkGreen}{\ding{51}} \\
\bottomrule
\end{tabular}%
}
\end{table}

\begin{figure*}[t]
    \centering
    \includegraphics[width=\textwidth]{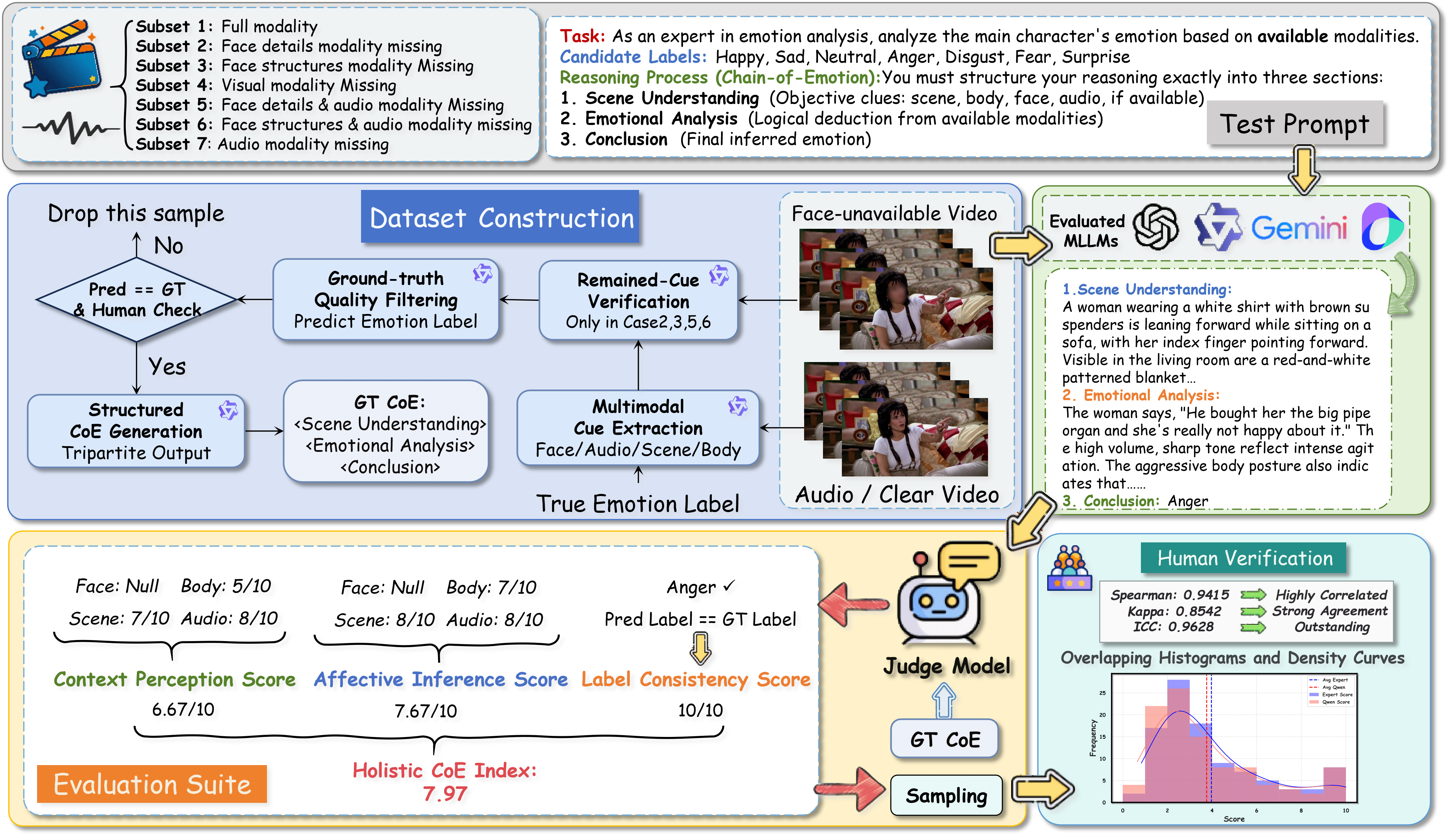}
    \caption{\textbf{Illustration of Our Data Construction and Evaluation Strategy.}}
    \Description{Overview of the MIME construction and evaluation pipeline, including modality degradation, cue verification, CoE generation, automatic scoring, and human validation.}
    \label{fig:pipeline}
\end{figure*}

\section{MIME Benchmark}

\subsection{Data Acquisition and Preprocessing}
To ensure diversity in social contexts and missing levels, we collect raw data from social media and four authoritative \textit{in-the-wild} benchmarks: \textbf{CAER}, \textbf{DFEW}, \textbf{MAFW}, and \textbf{FERV39k}. We focus on seven universal emotion categories: \textit{Happy, Sad, Neutral, Anger, Disgust, Fear,} and \textit{Surprise}. A balanced resampling strategy is applied to mitigate long-tail category bias, ensuring a rigorous evaluation of models' generalizability.

Through the modal degradation technique, we systematically decouple video signals into seven subsets to simulate real-world incompleteness:
Full modality (FM), Face details missing (FDM), Face structures missing (FSM), Visual modality missing (VMM), Face details and audio modality missing (FDAM), Face structures and audio modality missing (FSAM), Audio modality missing (AMM).

\subsection{Chain-of-Emotion (CoE) Generation}
To endow our dataset with rich, fine-grained reasoning annotations, we design a comprehensive four-stage pipeline utilizing \textbf{Qwen3-Omni} to generate high-quality Chain-of-Emotion. As shown in Fig.\ref{fig:pipeline}, this pipeline explicitly guides models to decouple and articulate their observations across individual modalities prior to predicting the final emotion.
The detailed information of the generation process can be found in the appendix.
The logic flow of CoE is divided into three folds:
\begin{itemize}
    \item \textbf{Scene Understanding ($C_{scene}$):} Characterizing the objective multimodal cues present in the input, extracting factual information exclusively from the available modalities (face, body, scene, and audio).
    \item \textbf{Emotional Analysis ($A_{emo}$):} Executing cognitive reasoning to map the previously identified cues to internal emotional states. This step explicitly articulates the logical path of how the observed signals lead to the emotional deduction.
    \item \textbf{Predicted Emotion ($L$):} Deriving the final discrete emotion category, which strictly aligns with the ground-truth label, based on the preceding systematic analysis.
\end{itemize}

\begin{figure}[t]
    \centering
    % 将 \linewidth 改为 0.85\linewidth 使其变小
    \includegraphics[width=0.9\linewidth]{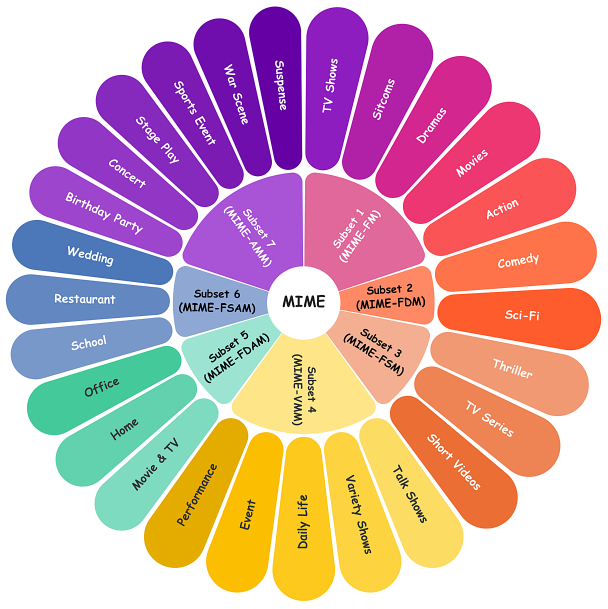}
    \caption{Overview of MIME's Task Types.Task types across diverse video scenarios.}
    \Description{MIME task types across diverse video scenarios.}
    \label{fig:intro1}
\end{figure}

\begin{figure}[t]
    \centering
    % 同样缩小为 0.85\linewidth
    \includegraphics[width=0.75\linewidth]{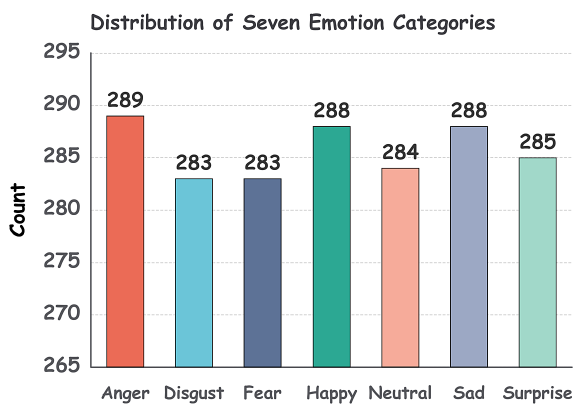}
    
    % \vspace{1em}
    
    \includegraphics[width=0.75\linewidth]{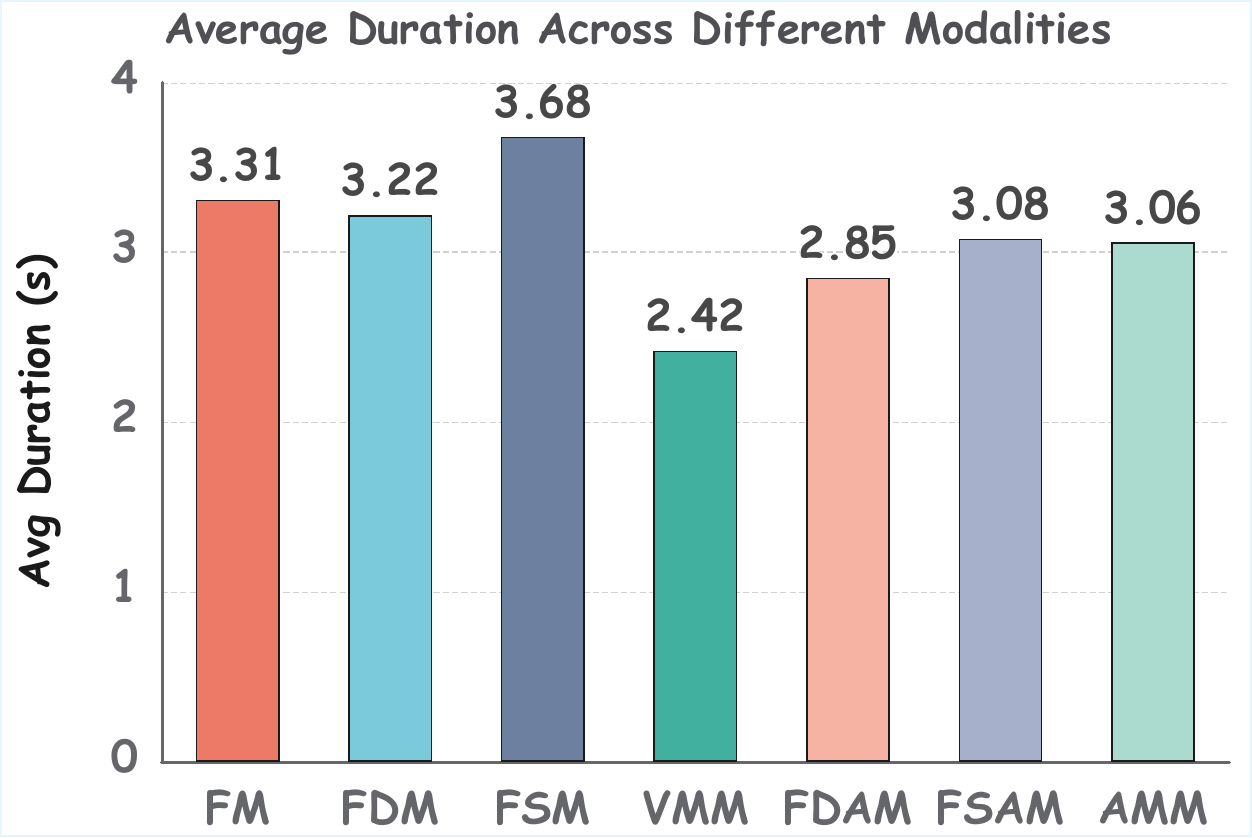}
    % \vspace{0.2em}
    \caption{Balanced distributions of emotion categories(top) and modality-specific video durations(bottom).}
    \Description{Distributions of emotion categories and average video durations across MIME subsets.}
    \label{fig:intro2}
\end{figure}

% \begin{figure}[h]
%     \centering
%     \includesvg[width=\linewidth]{figs/fig_radar.pdf}
%     \caption{\textbf{Performance Comparison of Leading MLLMs on \method{}.} Our evaluation suite assesses MLLMs using three unified metrics (Rec-S, Rea-S, and CoT-S) across eight emotional tasks.}
%     \label{fig:radar}
% \end{figure}

% \begin{figure}[h]
%     \centering
%     \includesvg[width=\linewidth]{figs/fig_main.pdf}
%     \caption{\textbf{Illustration of Our Evaluation Strategy.} We leverage a multi-agent system framework to assess the recognition and reasoning capabilities of MLLMs across different tasks with three unified metrics. To validate the effectiveness of our MLLM-as-judge strategy, we further compare the results of the judge agent on sampled data against results cross-evaluated by five human experts.}
%     \label{fig:eval_strategy}
% \end{figure}

% \begin{figure}[h]
%     \centering
%     \includesvg[width=\linewidth]{figs/fig_case_error.pdf}
%     \caption{\textbf{Cases of Error Responses.} We showcase some typical error types of cutting-edge MLLMs.}
%     \label{fig:case_error}
%     \vspace{-1em}
% \end{figure}

\subsection{Statistical Analysis}
To ensure the robustness and generalizability of the proposed benchmark, we conduct a comprehensive statistical analysis of the constructed MIME dataset, which contains 2,000 segments. Fig.\ref{fig:intro1} and Fig.\ref{fig:intro2} presents an \textbf{Overview of MIME Statistics}. As illustrated in the \textbf{Task Types (Fig.\ref{fig:intro1})}, the dataset exhibits a hierarchical distribution: it encompasses specific modality missingness types (inner circle, e.g., Audio Miss, Heavy degeneration Face, Visual Miss) systematically decoupled across diverse in-the-wild scene contexts (outer circle, e.g., Lifestyle, Movie, Vlog), enabling fine-grained analysis of diverse video contexts. Fig.\ref{fig:intro2}-Top demonstrates that the dataset rigorously adheres to a balanced sample distribution across the seven universal emotion categories, effectively mitigating the long-tail bias prevalent in in-the-wild datasets. Furthermore, as shown in the Fig.\ref{fig:intro2}-Bottom, the average durations of the audio/video data across different modality combinations are evenly distributed, facilitating a comprehensive evaluation of temporal understanding under various degraded cases.

\subsection{Evaluation Suite}
\subsubsection{Evaluation Strategy}
Traditional evaluation frameworks predominantly rely on discrete emotion labels, which struggle to comprehensively measure the reasoning quality of models in complex, modality-impaired scenarios. Inspired by the annotation-free evaluation paradigms introduced in recent emotional intelligence benchmarks \cite{zhang2025mme}, we introduce a hybrid evaluation strategy grounded in the LLM-as-a-Judge mechanism. Specifically, we deploy Qwen3.5-plus as a judge model to conduct automated and fine-grained quantitative scoring on the CoE generated by the evaluated models. This approach not only evaluates the final classification accuracy but also objectively measures the logical rigor and factual basis of the models' subjective reasoning processes, all while circumventing prohibitive manual annotation costs.
% \textbf{Figure 3} provides an \textbf{Illustration of Our Evaluation Strategy}. 
% Diverging from traditional evaluation frameworks that rely solely on classification accuracy, we introduce a novel Dual-Stage Hybrid Metrics system. This system is explicitly designed to quantify both the objective results and the reasoning process. 
\subsubsection{Evaluation Metrics}
To comprehensively dissect the emotional intelligence of MLLMs from both objective classification performance and subjective interpretability dimensions, we compartmentalize our evaluation metrics into the following \textbf{two stages}:
\paragraph{Stage I: Standard Classification metrics}
For the final emotion prediction, we compute the \textbf{Accuracy (Acc)} alongside \textbf{F1-score} to rigorously evaluate the models' classification robustness across the seven emotion categories, particularly mitigating potential long-tail distribution issues under modality degradation.

\paragraph{Stage II: CoE Quality Metrics}
To meticulously quantify the "compensatory reasoning" capability, the LLM-as-a-Judge conducts fine-grained scoring on the generated CoE based on four predefined modalities: Face ($f$), Body ($b$), Scene ($s$), and Audio ($a$). We denote the modality set as $\mathcal{M} = \{f, b, s, a\}$, and let $I_m \in \{0,1\}$ indicate whether modality $m$ is valid for scoring, where $I_m=1$ if modality $m$ is available or degraded-but-observable, and $I_m=0$ if modality $m$ is truly missing. The evaluation consists of \textbf{three sub-metrics} and \textbf{one holistic index}:

\textbf{Context Perception Score (CPS):} Evaluates whether the model accurately captures the surviving physical cues. The overall CPS is computed over the valid modalities only, where $s^{cps}_m \in [0,10]$:
% \[
% CPS = \sum_{m \in \mathcal{M}} w_m \cdot s^{cps}_m
% \]
\begin{equation}
CPS = \frac{\sum_{m \in \mathcal{M}} I_m \cdot s^{cps}_m}{\sum_{m \in \mathcal{M}} I_m}
\label{eq:cps}
\end{equation}

\textbf{Affective Inference Score (AIS):} Assesses the logical rigor of deducing emotional states from the perceived cues of each modality, explicitly penalizing hallucination, where each valid modality receives a score $s^{ais}_m \in [0,10]$:
% \[
% AIS = \sum_{m \in \mathcal{M}} w_m \cdot s^{ais}_m
% \]

\begin{equation}
AIS = \frac{\sum_{m \in \mathcal{M}} I_m \cdot s^{ais}_m}{\sum_{m \in \mathcal{M}} I_m}
\label{eq:ais}
\end{equation}

\textbf{Label Consistency Score (LCS):} Checks whether the semantic conclusion of the long-text reasoning path strictly aligns with the final categorical prediction. LCS equals 10 if correct, otherwise 0.

\textbf{Holistic CoE Index (HCI):} Serves as the ultimate benchmark for measuring the model's comprehensive "affective interpretability." It is the weighted integration of the above three metrics:
% $$HCI = \alpha \cdot CPS + \beta \cdot AIS + (1-\alpha-\beta) \cdot LCS$$
\begin{equation}
HCI = \alpha \cdot CPS + \beta \cdot AIS + (1-\alpha-\beta) \cdot LCS
\label{eq:hci}
\end{equation}

where the hyperparameters are empirically set as $\alpha = 0.4$ and $\beta = 0.3$ to emphasize foundational context understanding.

\subsubsection{Human Verification.}
% To validate the dependability of the proposed MLLM-as-judge framework, a manual evaluation was conducted by a panel of 3 experts. During this process, the evaluators annotated 173 distinct reasoning steps derived from a random subset of 100 questions and their corresponding CoE outputs. We subsequently analyzed the alignment between the Qwen-generated scores and those provided by the human experts.
% As illustrated in the bottom-right panel of Fig. \ref{fig:pipeline}, the Spearman's rank correlation coefficient \cite{spearman1961proof} comparing the two sets of evaluations is 0.9415, revealing a strong positive correlation. When assessing inter-rater reliability, the paired scores yield a Cohen’s Kappa coefficient \cite{mchugh2012interrater} of 0.8542, denoting near-perfect agreement. Moreover, an intra-class correlation coefficient (ICC) \cite{koch2004intraclass} of 0.9628 substantiates the outstanding consistency between the two grading approaches.
% Overall, these metrics confirm that our MLLM-as-judge methodology tightly mirrors human assessment, establishing it as a credible and robust evaluation approach.
To validate the reliability of our proposed MLLM-as-judge framework, we conducted manual evaluation with 3 expert annotators. 
The experts annotated 173 samples and corresponding CoE outputs to analyze the alignment between Qwen-generated scores and human annotations.
As shown in the bottom-right panel of Fig. \ref{fig:pipeline}, the two evaluations achieve a Spearman's rank correlation coefficient \cite{spearman1961proof} of 0.9415 (strong positive correlation), a Cohen’s Kappa coefficient \cite{mchugh2012interrater} of 0.8542 (near-perfect inter-rater agreement), and an intra-class correlation coefficient (ICC) \cite{koch2004intraclass} of 0.9628 (outstanding grading consistency).
These metrics confirm that our MLLM-as-judge methodology closely mirrors human assessment, verifying its credibility and robustness as an evaluation approach.

\section{Experiments}

\subsection{Experimental Setup}
We evaluate the performance on 7 cutting-edge MLLMs, including the GPT-5 series (gpt-5.4-nano, gpt-5-mini, gpt-5-nano), the Gemini series (gemini-3.1-flash), the Qwen suite (qwen3.5-plus, qwen3.5-27b), and doubao-seed-2-0. 
% For the open-sourced MLLMs (e.g., qwen3.5-27b), we use the vLLM framework to perform inference on our benchmark. For models that are not compatible with the vLLM framework, we conduct experiments using the official open-source code provided by the authors. 
% For closed-sourced MLLMs, inference is conducted directly via their official APIs. 
All the experiments are conducted under zero-shot settings. 
% To ensure a fair comparison across models, we maintain consistent prompting wherever possible, with only minor modifications when necessary (e.g., replacing "video" with "video frames" for vision-language models that do not support direct video input). 
Furthermore, to achieve scalable qualitative evaluation, we employ Qwen3.5-Plus as the primary LLM-as-a-Judge. Leveraging a concurrent execution architecture with multi-threading, the judge model scores the generated CoE texts based on their logical alignment and factual grounding.

\begin{table}[t]
\centering
\caption{\textbf{Overall Performance Comparison (\%) on \method{}.}
The top three performing results are highlighted in
\colorbox{backred!50}{red} (1\textsuperscript{st}),
\colorbox{backblue!75}{blue} (2\textsuperscript{nd}), and
\colorbox{lightyellow!40}{yellow} (3\textsuperscript{rd}).}
\label{tab:main_exp}
\resizebox{\columnwidth}{!}{%
\begin{tabular}{lccccc}
\toprule
\textbf{Model} & \textbf{F1}& \textbf{CPS} & \textbf{AIS} & \textbf{LCS} & \textbf{HCI} \\
\midrule
Qwen3.5-27b \cite{qwen3.5} 
& 40.99& \cellcolor{backblue!75}6.57 
& \cellcolor{lightyellow!40}5.13 
& 4.32 
& \cellcolor{lightyellow!40}5.46 \\
Qwen3.5-plus \cite{qwen3.5} 
& \cellcolor{lightyellow!40}42.60& \cellcolor{backred!50}\textbf{6.70} 
& \cellcolor{backred!50}\textbf{5.32} 
& \cellcolor{lightyellow!40}4.43 
& \cellcolor{backred!50}\textbf{5.60} \\
GPT-5.4-nano \cite{gpt54nano} 
& 26.21& 5.41 
& 3.97 
& 2.93 
& 4.23 \\
Gemini-3.1-flash \cite{gemini3} 
& \cellcolor{backred!50}\textbf{46.61}& \cellcolor{lightyellow!40}6.54 
& \cellcolor{backblue!75}5.24 
& \cellcolor{backred!50}\textbf{4.68} 
& \cellcolor{backblue!75}5.59 \\
GPT-5-mini \cite{gpt5mini} 
& 33.52& 5.91 
& 4.42 
& 3.48 
& 4.73 \\
Doubao-seed-2-0 \cite{seed20} 
& \cellcolor{backblue!75}45.27& 6.43 
& 4.96 
& \cellcolor{backblue!75}4.65 
& \cellcolor{lightyellow!40}5.46 \\
GPT-5-nano \cite{gpt5nano}& 27.01& 5.53 
& 4.06 
& 2.95 
& 4.31 \\
\bottomrule
\end{tabular}%
}
\end{table}

\begin{table*}[t]
\centering
\caption{Classification performance under different subsets.}
\label{tab:condition_results}
\resizebox{\textwidth}{!}{
\begin{tabular}{lcccccccccccccc}
\toprule
\multirow{2}{*}{\textbf{Model}}
& \multicolumn{2}{c}{\textbf{FM}}
& \multicolumn{2}{c}{\textbf{FDM}}
& \multicolumn{2}{c}{\textbf{FSM}}
& \multicolumn{2}{c}{\textbf{VMM}}
& \multicolumn{2}{c}{\textbf{FDAM}}
& \multicolumn{2}{c}{\textbf{FSAM}}
& \multicolumn{2}{c}{\textbf{AMM}} \\
\cmidrule(lr){2-3}\cmidrule(lr){4-5}\cmidrule(lr){6-7}\cmidrule(lr){8-9}\cmidrule(lr){10-11}\cmidrule(lr){12-13}\cmidrule(lr){14-15}
& \textbf{ACC} & \textbf{WF1} & \textbf{ACC} & \textbf{WF1} & \textbf{ACC} & \textbf{WF1} & \textbf{ACC} & \textbf{WF1} & \textbf{ACC} & \textbf{WF1} & \textbf{ACC} & \textbf{WF1} & \textbf{ACC} & \textbf{WF1} \\
\midrule
gpt-5.4-nano \cite{gpt54nano}
& 22.81 & 18.92
& 25.95 & 19.70
& 31.12 & 27.80
& 43.50 & 42.28
& 20.67 & 16.29
& 31.94 & 29.72
& 25.00 & 21.03 \\
gpt-5-mini \cite{gpt5mini}
& 27.07 & 24.13
& 24.68 & 22.80
& 36.93 & 34.70
& \textbf{50.00} & \textbf{48.78}
& 23.08 & 21.98
& 44.27 & 42.36
& 31.75 & 31.64 \\
gemini-3.1-flash \cite{gemini3}
& 46.10 & 44.83
& \textbf{50.96} & \textbf{49.82}
& \textbf{49.38} & \textbf{50.09}
& 44.25 & 42.75
& 37.98 & 37.31
& 48.42 & 47.33
& \textbf{50.75} & \textbf{51.34} \\
gpt-5-nano \cite{gpt5nano}
& 23.75 & 19.87
& 20.89 & 16.48
& 28.22 & 25.22
& 49.50 & 47.32
& 19.23 & 16.61
& 35.94 & 33.47
& 21.75 & 17.98 \\
qwen3.5-plus \cite{qwen3.5}
& 48.12 & 45.32
& 44.94 & 42.26
& 40.25 & 38.79
& 42.50 & 39.87
& 35.58 & 32.30
& \textbf{50.26} & \textbf{49.28}
& 46.00 & 44.44 \\
qwen3.5-27b \cite{qwen3.5}
& 44.11 & 40.57
& 48.73 & 46.81
& 42.74 & 40.79
& 42.61 & 39.48
& 35.10 & 30.75
& 46.35 & 44.43
& 43.50 & 40.64 \\
doubao-seed-2-0 \cite{seed20}
& \textbf{51.13} & \textbf{48.10}
& 48.73 & 48.29
& 48.13 & 45.61
& 41.50 & 39.24
& \textbf{40.87} & \textbf{38.02}
& 47.92 & 47.65
& 47.12 & 45.72 \\
\bottomrule
\end{tabular}
}
\end{table*}

%已核对
\begin{table*}[t]
\centering
\caption{Per-emotion Accuracy (ACC) and F1-score (F1) comparison of different models.}
\label{tab:emotion_f1}
\resizebox{\textwidth}{!}{
\begin{tabular}{l c c c c c c c c c c c c c c}
\toprule
\multirow{2}{*}{\textbf{Model}} & \multicolumn{2}{c}{\textbf{Happy}} & \multicolumn{2}{c}{\textbf{Sad}} & \multicolumn{2}{c}{\textbf{Neutral}} & \multicolumn{2}{c}{\textbf{Anger}} & \multicolumn{2}{c}{\textbf{Disgust}} & \multicolumn{2}{c}{\textbf{Fear}} & \multicolumn{2}{c}{\textbf{Surprise}} \\
\cmidrule(lr){2-3} \cmidrule(lr){4-5} \cmidrule(lr){6-7} \cmidrule(lr){8-9} \cmidrule(lr){10-11} \cmidrule(lr){12-13} \cmidrule(lr){14-15}
& ACC & F1 & ACC & F1 & ACC & F1 & ACC & F1 & ACC & F1 & ACC & F1 & ACC & F1 \\
\midrule
gpt-5.4-nano \cite{gpt54nano} & 15.28 & 0.23 & 26.04 & 0.27 & 33.10 & 0.28 & 53.50 & 0.41 & 9.19 & 0.16 & 29.72 & 0.35 & 8.42 & 0.14 \\
gpt-5-mini \cite{gpt5mini} & 21.18 & 0.28 & 33.68 & 0.36 & 51.41 & 0.31 & 57.14 & 0.43 & 11.66 & 0.20 & 45.23 & 0.45 & 23.51 & 0.30 \\
gemini-3.1-flash \cite{gemini3} & 50.69 & 0.57 & 53.31 & \textbf{0.52} & 68.20 & \textbf{0.41} & 56.14 & 0.52 & \textbf{21.20} & \textbf{0.32} & 37.81 & 0.46 & 40.14 & \textbf{0.47} \\
gpt-5-nano \cite{gpt5nano} & 12.15 & 0.19 & 24.31 & 0.29 & \textbf{68.66} & 0.31 & 35.07 & 0.39 & 4.59 & 0.09 & \textbf{45.58} & 0.38 & 16.49 & 0.24 \\
qwen3.5-plus \cite{qwen3.5} & 63.76 & \textbf{0.59} & \textbf{56.25} & 0.48 & 50.00 & 0.35 & 56.45 & \textbf{0.54} & 7.77 & 0.14 & 42.4 & \textbf{0.48} & 32.63 & 0.40 \\
qwen3.5-27b \cite{qwen3.5} & 66.67 & 0.58 & 55.56 & 0.45 & 43.66 & 0.35 & 50.87 & 0.49 & 6.71 & 0.12 & 42.05 & 0.45 & 35.92 & 0.43 \\
doubao-seed-2-0 \cite{seed20} & \textbf{68.75} & \textbf{0.59} & 48.61 & 0.50 & 57.24 & 0.40 & \textbf{57.84} & 0.53 & 17.31 & 0.27 & 31.45 & 0.41 & \textbf{43.51} & 0.46 \\
\bottomrule
\end{tabular}
}
\end{table*}

\subsection{Main Result}

We comprehensively showcase the performance comparison of various models on the MIME benchmark in Tab.\ref{tab:main_exp} and \ref{tab:condition_results}, encompassing both standard classification metrics (e.g., ACC and WF1) and CoE Quality Metrics scored by the judge agent.

\paragraph{Overall Classification and Reasoning Performance} The experimental landscape clearly reveals the immense challenge of multimodal emotion recognition under modality-impaired scenarios. Even the top-performing model, gemini-3.1-flash, achieves F1-Score of only 46.61 and an HCI of 5.59. It is closely followed by doubao-seed-2-0 (F1: 45.27, HCI: 5.46) and qwen3.5-plus (F1: 42.60, HCI: 5.60). In stark contrast, smaller-scale models such as gpt-5.4-nano and gpt-5-nano experience a precipitous performance drop, with F1-Score plummeting to 26.21 and 27.01, respectively, and their HCI hovering around 4.23 and 4.31. The fact that all evaluated models report F1-Score below 50 indicates that even state-of-the-art MLLMs are still in the nascent stages of developing robust emotional intelligence for complex, in-the-wild environments.

\paragraph{The Gap Between Context Perception and Affective Inference} Delving into the CoE quality metrics, we observe a pronounced divergence between "Scene Understanding (CPS)" and "Emotional Analysis (AIS)." For instance, while qwen3.5-plus achieves a high score of 6.70 in perceiving foundational physical cues (CPS), its score drastically drops to 5.32 when translating these cues into internal affective inference (AIS). Similarly, gpt-5-mini scores 5.91 and 4.42 on these two metrics, respectively. This phenomenon profoundly exposes the affective cognitive bottleneck of current MLLMs: they are capable of adequately "seeing" and extracting objective physical elements in degraded environments, but they severely lack the capacity to execute deep, compensatory logical reasoning based on those observations.

\paragraph{Modality Contribution under Incomplete Conditions}
Table~\ref{tab:condition_results} shows the effect of different modalities on emotion recognition. Comparing FDM with FDAM highlights audio contribution: with degraded facial details, removing audio causes clear WF1 drops, such as qwen3.5-27b from 46.81 to 30.75 (-16.06) and gemini-3.1-flash from 49.82 to 37.31 (-12.51). This indicates that audio compensates when facial information is unreliable.

\subsection{Observations and Insights}

Through cross-analysis of the quantitative classification results and qualitative CoE evaluations, we distill three insights regarding MLLMs' affective reasoning under modality-impaired conditions (Subsets 1--7):

\paragraph{Obs. 1: A Pronounced Gap Between Visual Perception and Cognitive Reasoning} "Seeing" a scene does not guarantee "understanding" its emotion. Across all models, CPS (Context Perception Score) consistently exceeds AIS (Affective Inference Score). For instance, qwen3.5-plus accurately extracts physical cues (CPS: 6.70/10), but struggles to logically deduce internal emotional states (AIS: 5.32/10). This indicates that while MLLMs possess strong visual foundations, they lack human-like empathic cognition to bridge the semantic gap from objective physical space to subjective affective space.

\paragraph{Obs. 2: Robust IMER Requires Cross-Modal Compensation Rather Than Single-Cue Recognition} The subset-level results indicate that incomplete emotion recognition cannot rely on a single dominant cue. Facial appearance and audio provide direct emotional evidence, while body posture and scene context serve as compensatory cues. However, current MLLMs still struggle to dynamically reweight available cues. When facial details and audio are jointly unavailable, models infer emotions mainly from body language and scene context, leading to clear degradation. Thus, the key challenge of IMER is adaptively shifting attention among remaining modalities under different missing conditions.
% \textbf{Obs. 3: CoE Quality Closely Tracks Prediction Reliability.} We observe a strong positive correlation between classification accuracy and CoE quality measured by HCI. Stronger models (e.g., gemini-3.1-flash, qwen3.5-plus, doubao-seed-2-0) maintain HCI scores above 5.45, while lightweight GPT-nano models stay below 4.40 and produce fragmented reasoning traces. This suggests that better CoE quality serves as a reliability anchor under incomplete conditions. In other words, models that generate more coherent and evidence-grounded reasoning are also more likely to remain robust when dominant modalities are missing.

\paragraph{Obs. 3: Explicit CoE Quality Dictates Predictions in Complex Scenarios}
Our analysis shows a strong positive correlation between classification accuracy and CoE reasoning quality, measured by HCI. State-of-the-art models (e.g., gemini-3.1-flash, qwen3.5-plus) consistently achieve HCI scores above 5.45, whereas lightweight GPT-nano models, with F1-scores below 30, produce fragmented reasoning traces and score under 4.40. Leading models also obtain higher Label Consistency Scores (LCS), suggesting that their predictions rely on stable step-wise deduction rather than heuristic shortcuts. Thus, high-quality CoE reasoning serves as a reliability anchor under incomplete or degraded multimodal cues, making the reinforcement of MLLMs' internal ``Compensatory Chain-of-Emotion'' a promising direction for mitigating multimodal affective bottlenecks.% Through a rigorous analysis of the performance of various MLLMs on MIME, we summarize several key observations that highlight the current limitations of emotional intelligence under modality-compromised scenarios.

\begin{figure}[t]
    \centering
    \includegraphics[width=0.92\columnwidth]{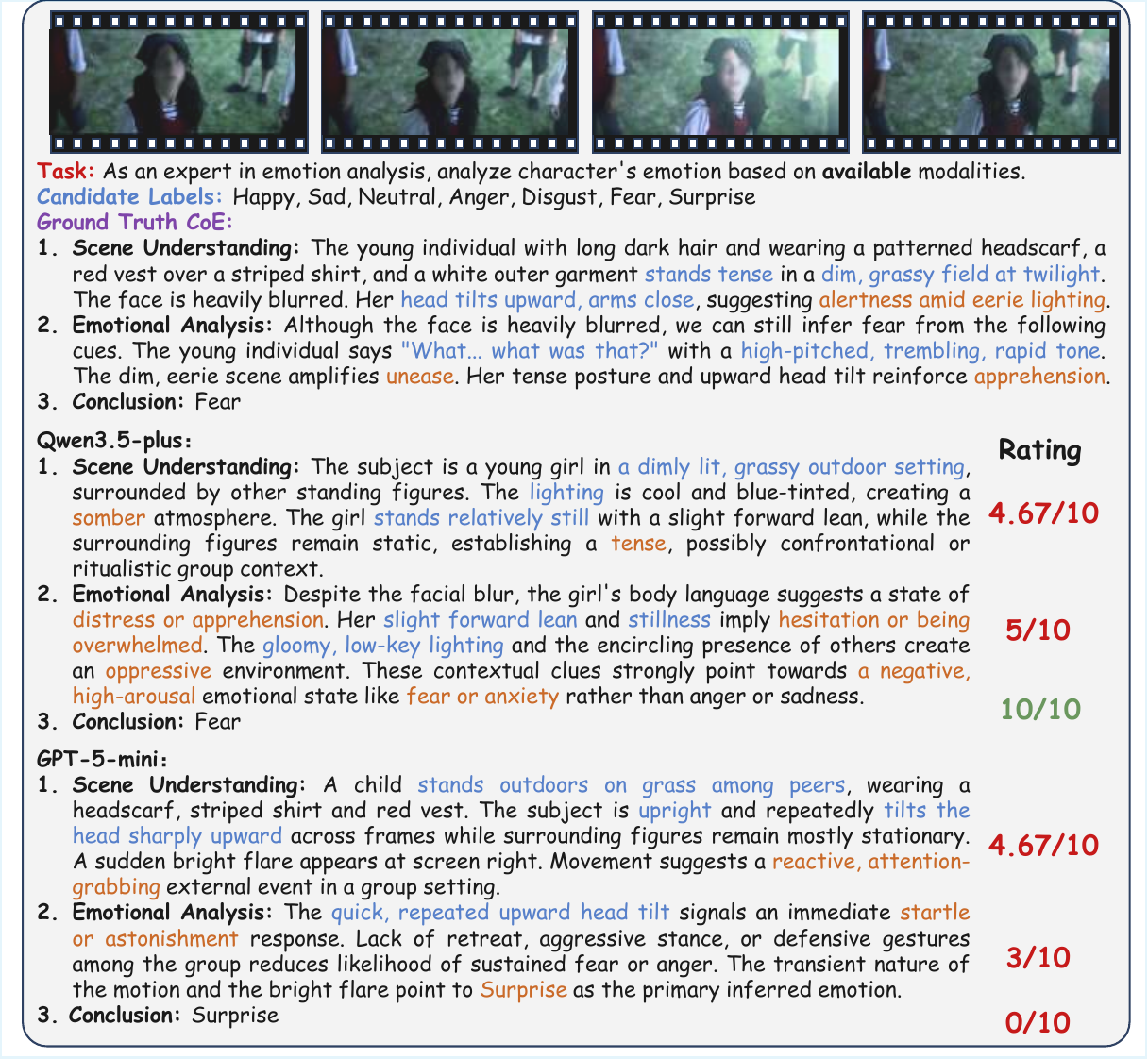}
    \caption{\textbf{Error cases of model responses.}}
    \Description{Examples comparing ground-truth CoE reasoning with model responses under degraded multimodal cues.}
    \label{fig:failed_example}
\end{figure}

% \subsection{Qualitative Case Study}
% To intuitively illustrate the specific vulnerabilities of current models, \textbf{Figure 4} showcases \textbf{Cases of Error Responses}, highlighting typical error types of cutting-edge MLLMs. When faced with degraded or complex emotional contexts, models such as Video-LLaMA2, Qwen2.5-Omni, and Gemini-2.5-Pro exhibit significant limitations in affective interpretability. For instance, when facial features or audio are obscured, models easily overlook crucial secondary physical cues. In many cases, instead of executing genuine compensatory reasoning, the models suffer from severe hallucination, tending to fabricate non-existent facial expressions or simply outputting neutral predictions because the dominant cues are missing. This intuitive comparison vividly highlights the immense potential of our benchmark in guiding large models to learn human-level robust reasoning mechanisms. 

\section{Conclusion}

% In this paper, we tackled a critical yet underexplored challenge in affective computing: multimodal emotion recognition under missing modalities. Recognizing the limitations of existing datasets that assume ideal full-modality inputs and provide only discrete labels, we introduced MIME, the first large-scale Chain-of-Emotion benchmark. By systematically simulating real-world modality degradation across seven distinct scenarios—including progressive dynamic face-degeneration and audio removal—we established a rigorous testing environment. Driven by an adaptive four-stage generation pipeline with a conditional hallucination-suppression mechanism, our dataset provides high-quality, tripartite structured reasoning chains. Furthermore, we designed a dual-stage hybrid evaluation suite, utilizing an LLM-as-a-Judge to comprehensively quantify both objective classification robustness and subjective reasoning logic. Extensive benchmarking of state-of-the-art MLLMs reveals their significant vulnerabilities under modality absence. We hope MIME will serve as a foundational stepping stone, inspiring future research to break the boundaries of facial reliance and foster highly robust, interpretable emotional intelligence in open-world scenarios.

In this paper, we address a critical underexplored challenge in affective computing: multimodal emotion recognition under missing modalities. We further propose MIME, the first fine-grained Chain-of-Emotion benchmark. It systematically simulates real-world modality degradation across seven scenarios to build a rigorous testing environment. Built via an adaptive four-stage generation pipeline with conditional hallucination suppression, MIME delivers high-quality tripartite structured reasoning chains. We also design a dual-stage hybrid evaluation suite with an LLM-as-a-Judge, to quantify models' objective classification robustness and subjective reasoning logic.
% Extensive benchmarking on state-of-the-art MLLMs reveals their significant vulnerabilities under modality absence. 
% We expect MIME to serve as a foundational benchmark, inspiring research to break facial reliance and advance robust, interpretable open-world emotional intelligence.

\begin{acks}
    This work was supported by the National Natural Science Foundation of China (Grant No. 62576085, 62236006, 62320106007 and 62521007), the Fundamental Research Funds for the Central Universities (Grant No. N25XQD057), and the National Postdoctoral Researcher Program of China, Category C (Grant No. GZC20260869). This work was supported by the Open Research Fund of the State Key Laboratory of Novel Software Technology, Nanjing University (Grant No. KFKT2026B56). This work was supported by the Open Project of the Hubei Key Laboratory of Multimedia and Network Communication Engineering. This work was supported by the National Engineering Research Centre for Multimedia Software, Wuhan University. This work is supported by Open Research Fund of The State Key Laboratory of Multimodal Artificial Intelligence Systems (Grant No. MAIS2026081). This work was supported by Liaoning Collaboration Innovation Center For CSLE.
\end{acks}

% \section*{Acknowledgments}
% This work was supported by the National Natural Science Foundation of China (Grant No. 62576085), the Fundamental Research Funds for the Central Universities (Grant No. N25XQD057), and the National Postdoctoral Researcher Program of China, Category C (Grant No. GZC20260869). This work was supported by the Open Research Fund of the State Key Laboratory of Novel Software Technology, Nanjing University (Grant No. KFKT2026B56). This work was supported by the Open Project of the Hubei Key Laboratory of Multimedia and Network Communication Engineering. This work was supported by the National Engineering Research Centre for Multimedia Software, Wuhan University. This work is supported by Open Research Fund of The State Key Laboratory of Multimodal Artificial Intelligence Systems (Grant No. MAIS2026081). This work was supported by Liaoning Collaboration Innovation Center For CSLE.

\bibliographystyle{ACM-Reference-Format}
\balance
\bibliography{main}
\end{document}